\documentclass[a4paper]{jpconf}
\usepackage{graphicx}
\usepackage{dcolumn}
\usepackage{bm}
\usepackage{color}
\usepackage{ulem}
\usepackage{amsmath}

\begin{document}
\title{Toroidal order in a partially disordered state on a layered triangular lattice: 
implication to UNi$_4$B}

\author{Satoru Hayami$^1$, Hiroaki Kusunose$^2$, and Yukitoshi Motome$^1$}

\address{$^1$ Department of Applied Physics, University of Tokyo, Tokyo 113-8656, Japan}
\address{$^2$ Department of Physics, Ehime University, Matsuyama 790-8577, Japan
}

\ead{hayami@aion.t.u-tokyo.ac.jp}

\begin{abstract}
A partial disorder on a layered triangular lattice is theoretically investigated from a viewpoint of toroidal ordering and magnetoelectric effects. 
We consider an extended periodic Anderson model including a site-dependent antisymmetric spin-orbit coupling between conduction and localized electrons. 
We show that, by the mean-field approximation, the model exhibits a coplanar vortex-lattice-type magnetic order as observed in a hexagonal uranium compound UNi$_4$B, in the parameter region with intermediate hybridization and electron correlation. 
This peculiar state accommodates a toroidal order, which leads to the linear magnetoelectric effect. 
We discuss the implications of our results to UNi$_4$B, focusing on the possible source of the site-dependent antisymmetric spin-orbit coupling. 
\end{abstract}

\section{Introduction}
Toroidal order is a peculiar magnetic state, which accompanies spontaneous breaking of both time-reversal and spatial-inversion symmetries. 
The toroidal moment is defined by the sum of outer products of magnetic moments and polarization vectors; 
its simplest form is given by spin ``loops" around the inversion centers. 
Recently, toroidal order has gained interest because it gives rise to cross correlations between electric and magnetic responses, such as magnetoelectric effects, diamagnetic anomaly, and nonreciprocal directional dichroism~\cite{gorbatsevich1994toroidal,popov1998magnetoelectric,schmid2001ferrotoroidics,EdererPhysRevB.76.214404,Spaldin:0953-8984-20-43-434203,kopaev2009toroidal}. 
Such experimental studies, however, are thus far restricted to insulating materials~\cite{FolenPhysRevLett.6.607,popov1999magnetic,arima2005resonant,MiyaharaJPSJ.81.023712,van2007observation,ToledanoPhysRevB.84.094421}.

Recent theoretical studies have cast a new light on the toroidal ordering from two viewpoints~\cite{Yanase:JPSJ.83.014703,Hayami_PhysRevB.90.024432}. 
One is the possibility of the toroidal order in metallic systems despite the absence of macroscopic polarization. 
The other is the effect of a site-dependent antisymmetric spin-orbit coupling, which appears in the lattice preserving the spatial-inversion (parity) symmetry globally but breaking intrinsically at each magnetic site. 
The representative lattice structures with such local parity breaking are zigzag chain, honeycomb lattice, and diamond lattice. 
In these systems, peculiar transport and magnetoelectric effects were predicted: 
for instance, a new-type of magnetotransport effect, that is, the intrinsic off-diagonal response without an external magnetic field~\cite{Hayami_PhysRevB.90.024432}.
In order to further stimulate experiments, it is desired to systematically study when and how such toroidal ordered metals are realized from the microscopic point of view. 

In the present study, we discuss the possibility of a metallic toroidal order in $f$ electron systems. 
We here focus on a uranium compound UNi$_4$B~\cite{Mentink1994,Oyamada2007,Haga,Oyamada2008-2} as a candidate material showing the toroidal nature. 
UNi$_4$B has been studied for its peculiar magnetic ordering, that is, a partial disorder, which is given by a periodic array of magnetically-ordered and nonmagnetic sites~\cite{Mekata_JPSJ.42.76}. 
The compound shows a second-order phase transition at 20~K to the partially disordered state. 
The magnetic unit cell consists of nine U sites: 
six out of the nine U sites develop magnetic moments and form a coplanar vortex-lattice-type order composed of hexagonal loops of the magnetic moments, while the rest three, each of which is surrounded by the magnetic U sites, remain nonmagnetic. 
The peculiar magnetic state was theoretically studied on the basis of a classical pseudospin model by the mean-field approximation~\cite{LacroixUNi4B}. 
However, the model was an effective localized spin model obtained by integrating out the interplay between conduction and localized electrons. 
Hence, it is not appropriate for describing the nature specific to metallic systems that we are interested in here. 
Moreover, despite the vortex-lattice-type order that consists of spin loops, a viewpoint of the toroidal ordering has been lacked in the previous studies. 
In particular, the role of the antisymmetric spin-orbit coupling has been fully neglected. 
In the following, we revisit this partial disorder with paying attention to the local parity breaking and toroidal ordering.

\section{Model}
\label{sec:Model}

In order to investigate the toroidal ordering in $f$ electron systems, we 
begin with the periodic Anderson model, which is a standard microscopic model describing the hybridization between conduction and localized electrons.  
We here consider an extension of the model to incorporate the effect of the antisymmetric spin-orbit coupling originating from the lattice symmetry. 
Specifically, we consider a layered triangular lattice, 
and take into account an antisymmetric hybridization between conduction and localized electrons, with keeping UNi$_4$B in mind. 
The Hamiltonian for the extended periodic Anderson model is given by 
\begin{align}
\label{eq:Ham_PAMSO_sec2}
\mathcal{H} 
= &-t \sum_{\langle i, j \rangle,\sigma} 
 (c^{\dagger}_{i \sigma} c_{j \sigma}   + {\rm H.c.} )- V  \sum_{i ,\sigma} 
( c^{\dagger}_{i \sigma}f_{i \sigma}+{\mathrm{H.c.}} )  
+ E_0 \sum_{i, \sigma} n_{i \sigma}^f \nonumber \\
&
+ U \sum_i n_{i \uparrow}^f n_{i \downarrow}^f 
+ \sum_i (\bm{s}^{cf}_i \times \bm{D}^{cf}_i)^z, 
\end{align}
where $c^{\dagger}_{i \sigma}$($c_{i \sigma}$) and $f^{\dagger}_{i \sigma}$($f_{i \sigma}$) 
are the creation (annihilation) operators of conduction and localized 
electrons with spin $\sigma$ at site $i$, and $n_{i\sigma}^{f} = f_{i\sigma}^\dagger f_{i\sigma}$. 
The first four terms in Eq.~(\ref{eq:Ham_PAMSO_sec2}) represent the standard periodic Anderson model: the kinetic energy of conduction electrons, the on-site hybridization between conduction and localized 
electrons, the atomic energy of $f$ electrons, and the on-site Coulomb interaction for $f$ electrons. 
The sum of $\langle i, j \rangle$ in the first term is taken over the nearest-neighbor sites on the layered triangular lattice. 
For simplicity, we assume the same transfer integral $t$ for intra- and inter-layer bonds; the results are not qualitatively altered for small anisotropy. 
Hereafter, we set $t$ as an energy unit and both the lattice constants in the intra- and inter-layer directions as a length unit.  
The last term is the antisymmetric hybridization term between conduction and localized electrons; $\bm{s}_i^{cf} = \sum_{\sigma \sigma'} ( c^{\dagger}_{i \sigma} \bm{\sigma}_{\sigma \sigma'} f_{i\sigma'} + {\rm H.c.})$. 
Here, we assume that the vector $\bm{D}^{cf}_i$ has site (sublattice) dependence, which characterizes the local parity breaking at each site, and 
originates from the odd-parity crystalline electric field discussed below. 
In the following calculations, we fix the electron density at half filling, $n=\sum_{i \sigma} \langle c_{i \sigma}^{\dagger}c_{i\sigma} + f_{i \sigma}^{\dagger}f_{i \sigma} \rangle/N=2$, where $N$ is the total number of sites.

\begin{figure}[htb!]
\begin{center}
\includegraphics[width=1.0 \hsize]{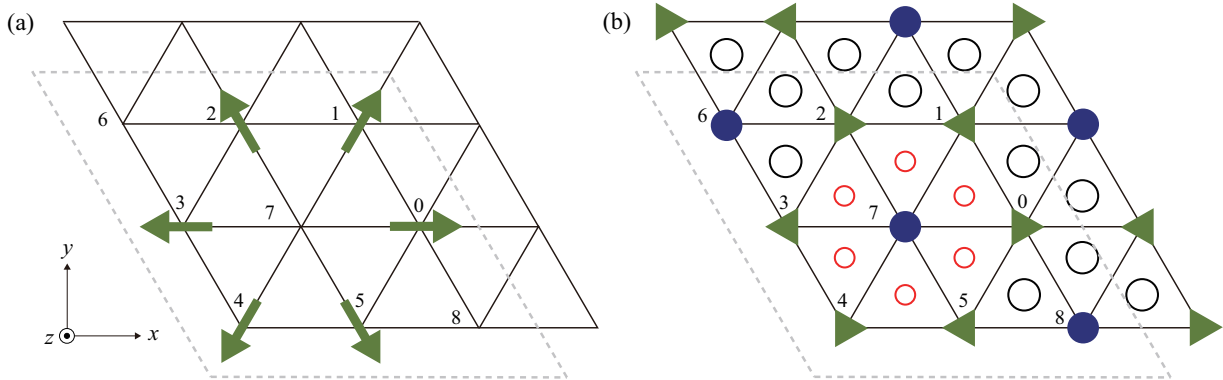} 
\caption{
\label{Fig:triangular_localinversion}
(a) Schematic picture of a triangular plane in the layered structure for the model in Eq.~(\ref{eq:Ham_PAMSO_sec2}). 
The green arrows indicate the directions of $\bm{D}^{cf}_{l}$ in Eq.~(\ref{eq:D-UNi4B}). 
The dashed diamond indicates the magnetic unit cell, and the numbers from $0$ to $8$ represent the sublattices. 
(b) Schematic picture of a model for UNi$_4$B, including 
two different types of nonmagnetic ions denoted by open blue and red circles at the centers of U triangular plaquettes. 
The configuration of the nonmagnetic ions is a simplified model for the experimental lattice structure in Ref.~\cite{Haga}. 
Filled blue circles and green triangles represent the crystallographically different lattice sites on the triangular plane, distinguished by the surrounding nonmagnetic ions. 
}
\end{center}
\end{figure}

In the present model, we assume the direction of $\bm{D}^{cf}_i$ in a nine-sublattice form with the specific directions in each triangular plane ($xy$ plane). 
The directions are assumed to be common among the triangular planes. 
The nine-sublattice form is represented by 
\begin{equation}
\label{eq:D-UNi4B}
\bm{D}^{cf}_l =
\begin{cases}
(\cos \frac{\pi}{3}l, \sin \frac{\pi}{3}l, 0) D\sin k_z  & (l=0,1,\dots, 5), \\
  0 & (l=6,7,8), 
\end{cases}
\end{equation}
where $l$ is the sublattice index [see Fig.~\ref{Fig:triangular_localinversion}(a)], and $D$ is a parameter to control the magnitude of the antisymmetric hybridization. 
The possible origins for $\bm{D}_i^{cf}$ in the nine-sublattice form are discussed in relation with UNi$_4$B in Sec.~\ref{sec:Discussion}.

\section{Result}

\begin{figure}[htb!]
\begin{center}
\includegraphics[width=1.0 \hsize]{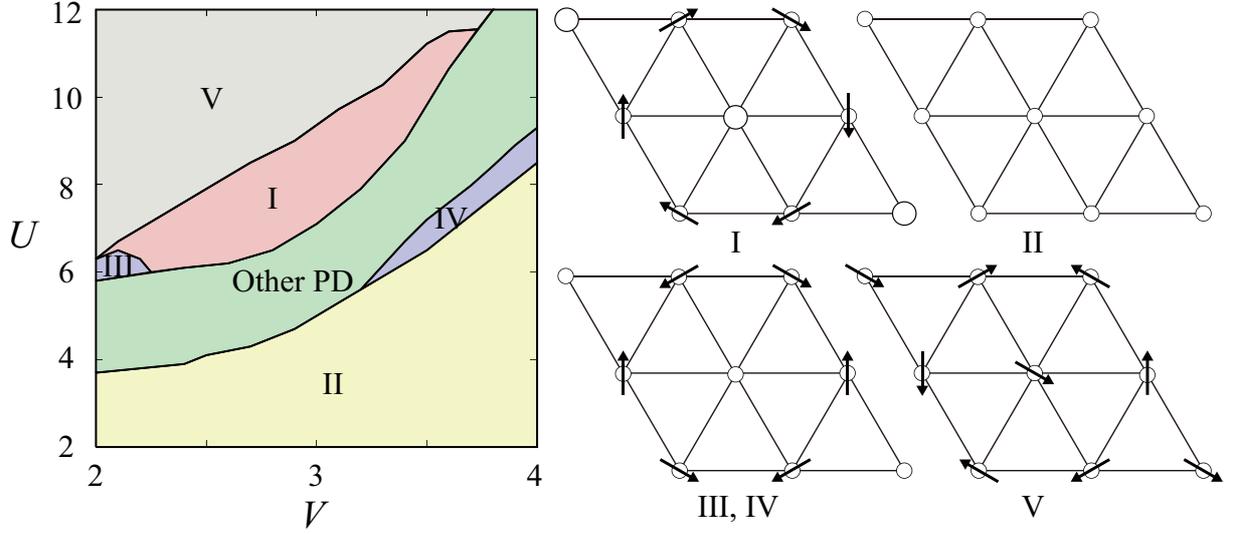} 
\caption{
\label{Fig:souzu_UNi4B}
Ground-state phase diagram of the model in Eq.~(\ref{eq:Ham_PAMSO_sec2}) with Eq.~(\ref{eq:D-UNi4B}) on a layered triangular lattice obtained by the mean-field calculations. 
The data are taken at half filling for $D=3$ and $E_0 = -6$. 
Schematic pictures of the ordering patterns are shown in the right panel. 
The arrows and the size of circles represent magnetic moments and local electron densities of $f$ electrons, respectively. 
The phases I, III, and IV are partially disordered states, while the phase V is a noncollinear antiferromagnetic phase. 
``Other PD" represents a complicated magnetically ordered state with partial disorder. 
See the text for details. 
}
\end{center}
\end{figure}

We study the ground state of the model in Eq.~(\ref{eq:Ham_PAMSO_sec2}) with Eq.~(\ref{eq:D-UNi4B}) by the standard Hartree-Fock approximation to the Coulomb $U$ term. 
We assume the same nine-sublattice form for the mean fields as that for $\bm{D}^{cf}_i$, but allow an arbitrary magnetic and charge pattern within the nine-site unit cell, similar to the method used in Ref.~\cite{Hayami_PhysRevB.90.024432}. 
We also assume that the magnetic moments are within the triangular planes. 
We calculate the mean fields by taking the sum over $12 \times 12 \times 36$ grid points in the folded Brillouin zone. 

Figure~\ref{Fig:souzu_UNi4B} shows the ground-state phase diagram at half filling obtained by the mean-field calculations. 
Schematic pictures of magnetic and charge states of $f$ electrons are shown in the right panel of Fig.~\ref{Fig:souzu_UNi4B}. 
The result shows that the system exhibits a variety of magnetic orders while changing the Coulomb interaction $U$ and the hybridization $V$. 

The most interesting phase is the partially disordered phase with a vortex-lattice-type magnetic structure, which is denoted by the phase I in Fig.~\ref{Fig:souzu_UNi4B}. 
This magnetic pattern coincides with that observed in UNi$_4$B~\cite{Mentink1994}. 
This phase I is stabilized in the region for intermediate $U$ and $V$. 
The system is metallic and shows charge disproportionation: the local charge density is higher at the nonmagnetic sites than the magnetic sites, as shown in the schematic picture in Fig.~\ref{Fig:souzu_UNi4B}. 
The tendency of charge disproportionation in the partially disordered state is the same as that in the partially disordered state in the periodic Anderson model without the antisymmetric hybridization, the last term in Eq.~(\ref{eq:Ham_PAMSO_sec2})~\cite{Hayami2011,Hayami2012} as well as in the Kondo lattice model~\cite{Motome2010}. 

This partially disordered state accommodates a toroidal order: the magnetic moments are perpendicular to $\bm{D}^{cf}_i$ at each site with forming spin loops around the nonmagnetic sites (sublattice 7 in Fig.~\ref{Fig:triangular_localinversion}). 
Indeed, this metallic phase exhibits a characteristic modulation of the band structure with a band bottom shift, anisotropic magnetotransport, and magnetoelectric effects, similar to the results in the toroidal ordered state discussed in Ref.~\cite{Hayami_PhysRevB.90.024432}. 
Such interesting aspects of the partially disordered state, however, have not been studied in experiments thus far. 
It is highly desired to examine this peculiar state experimentally from the viewpoint of toroidal ordering. 

Let us mention other phases appearing in Fig.~\ref{Fig:souzu_UNi4B}. 
The phase II in the small $U$ region is the nonmagnetic phase. 
This is a band insulator, whose gap is opened by the hybridization $V$. 
The phase III and IV are the partially disordered phases, whose magnetic structures 
are similar to each other but different from the phase I, as shown in Fig.~\ref{Fig:souzu_UNi4B}. 
The phase III is metallic with small charge disproportionation similar to that in the phase I, whereas the phase IV is insulating without any charge modulation. 
The phase V in the large $U$ region is the metallic antiferromagnetic phase, in which all the sites retain nonzero magnetic moments forming a noncollinear 120$^\circ$ structure, as shown in Fig.~\ref{Fig:souzu_UNi4B}. 

\section{Discussion}
\label{sec:Discussion}

In the previous section, we found the partially disordered phase showing the toroidal order in the model in Eq.~(\ref{eq:Ham_PAMSO_sec2}) with Eq.~(\ref{eq:D-UNi4B}). 
Obviously, the form of $\bm{D}^{cf}_i$ plays an important role in stabilizing this peculiar state. 
Let us here discuss the reason why we assume $\bm{D}^{cf}_i$ in the nine-sublattice form in Eq.~(\ref{eq:D-UNi4B}). 
In the assumption, we refer to the recent crystal structure data for UNi$_4$B obtained by the neutron diffraction experiment~\cite{Haga}. 
The results indicate that the nonmagnetic ions, Ni and B, are not either uniformly or randomly distributed in the U triangular lattice layers, but comprise a periodic structure. 
There, 1/3 of U sites are surrounded by the same ions (namely, all B or all Ni), whereas the rest 2/3 are surrounded by a mixture of B and Ni (two B and four Ni or four B and two Ni). 
This implies that the 1/3 U sites feel a rather symmetric crystalline electric field from the neighboring nonmagnetic ions, whereas the rest 2/3 are subject to a rather asymmetric one. 
Taking account of the characteristic of the local parity breaking, we mimic this experimental situation by the nine-sublattice form of $\bm{D}^{cf}_i$. 
Although the actual structure is more complicated with a larger unit cell~\cite{Haga}, 
we simplify the situation by assuming that two kinds of nonmagnetic ions, corresponding to B and Ni, are aligned eighteen-site unit cell, as denoted by open red and black circles in Fig.~\ref{Fig:triangular_localinversion}(b). 
This assumption leads to the nine-site sublattice of the triangular-lattice sites corresponding to U by differentiating two crystallographically different sites, as shown in Fig.~\ref{Fig:triangular_localinversion}(b): the 1/3 sites denoted by the filled blue circles feel a symmetric crystalline electric field, while the rest 2/3 denoted by the green triangles are subject to the asymmetric field in the direction connecting the site with the sublattice site $l=7$. 
The asymmetric crystalline electric field results in the local parity breaking at the sublattice sites $l=0, 1, \dots,$ and $5$. 
The principal direction of the asymmetry determines the direction of $\bm{D}^{cf}_i$ as Eq.~(\ref{eq:D-UNi4B}). 

Besides the structural origin from the nonmagnetic ions, the site-dependent anisotropic $\bm{D}^{cf}_i$ might also originate from the magnetostriction. 
In general, the bond length depends on both 
the relative angle between the magnetic moments at the both ends as well as their magnitudes. 
Hence, the U-U distances can be modulated according to the peculiar spin configuration in the partially disordered state below the transition temperature: 
by the symmetry argument, the magnetostriction arising from the vortex-lattice-type magnetic pattern leads to the same form of $\bm{D}^{cf}_i$ as in Eq.~(\ref{eq:D-UNi4B}). 
We note that this contribution appears only below the magnetic transition temperature. 
Although we assumed the model-embedded (temperature-independent) $\bm{D}^{cf}_i$ in the present analysis, a rather different situation may occur if $\bm{D}^{cf}_i$ becomes nonzero only by the magnetostriction, i.e., only once the partial disorder sets in. 
In this situation, the magnetoelectric effects appear only below the critical temperature.

\section{Summary}
We have examined the possibility of active toroidal moments in the partially disordered state, bearing UNi$_4$B in mind. 
Analyzing an extended periodic Anderson model including the antisymmetric hybridization between conduction and localized electrons, we have shown that the model exhibits a partial disorder with vortex-lattice-type magnetic ordering, similar to that in UNi$_4$B. 
Our results imply that the partially disordered state in UNi$_4$B may host a toroidal order. 
Hence, we anticipate that UNi$_4$B shows the interesting band deformation, magnetotransport, magnetoelectric effects, and intrinsic off-diagonal magnetoelectric response without an external magnetic field, as pointed out in Ref.~\cite{Hayami_PhysRevB.90.024432}. 

On the other hand, further analyses are desirable in order to test the validity of our assumption on the form of the antisymmetric hybridization. 
We have discussed two scenarios: the surrounding nonmagnetic sites and the magnetostriction. 
It might be possible to identify the origin of the antisymmetric hybridization experimentally, for instance, by the magnetoelectric response. 
In the former scenario, the staggered (toroidal) response is nonzero above the critical temperature and shows a peak near the critical temperature~\cite{Hayami_PhysRevB.90.024432}, while it becomes nonzero only below the critical temperature in the latter~\cite{Hayami_PhysRevB.90.081115}.

\ack

The authors thank Y. Haga and H. Amitsuka for fruitful discussions. 
S.H. is supported by Grant-in-Aid for JSPS Fellow. 
This work was supported by Grants-in-Aid for Scientific Research (No.~24340076), the Strategic Programs for Innovative Research (SPIRE), MEXT, and the Computational Materials Science Initiative (CMSI), Japan.

\section*{References}
\bibliographystyle{iopart-num}
\bibliography{ref}

\end{document}